\documentclass[11pt]{article}
\usepackage{amsmath,amssymb,graphicx}
\usepackage{hyperref}
\usepackage{cite}

\usepackage{tikz}
\usepackage{multicol,color,xcolor,longtable}

\definecolor{darkred}{rgb}{0.65,0.15,0}
\hypersetup{pdfborder={0 0 0},colorlinks=true,urlcolor=blue,citecolor=blue,linkcolor=darkred,linktocpage=true}

\newcommand{\nn}{\nonumber}
\newcommand{\cL}{\mathcal{L}}

\newcommand{\eprint}[1]{{\href{http://arxiv.org/abs/#1}{[\texttt{#1}]}}}
\newcommand{\eprintN}[1]{{\href{http://arxiv.org/abs/#1}{[\texttt{#1 [hep-th]}]}}}
\newcommand{\eprintGR}[1]{{\href{http://arxiv.org/abs/#1}{[\texttt{#1 [gr-qc]}]}}}
\newcommand{\eprintDG}[1]{{\href{http://arxiv.org/abs/#1}{[\texttt{#1 [math.DG]}]}}}
\newcommand{\doi}[2]{{\hypersetup{urlcolor=darkred}\href{http://dx.doi.org/#2}{#1}\hypersetup{urlcolor=blue}}}
\newcommand{\eprintCM}[1]{{\href{http://arxiv.org/abs/#1}{[\texttt{#1 [cond-mat.mes-hall]}]}}}
\newcommand{\be}{\begin{equation}}
\newcommand{\ee}{\end{equation}}
\newcommand{\bea}{\begin{eqnarray}}
\newcommand{\eea}{\end{eqnarray}}

\makeatletter

\@addtoreset{equation}{section}
\makeatother

\begin{document}

\mbox{}
\vspace{15mm}

\begin{center}
{\bf \Large Non-Lorentzian
theories with and without constraints}\\[12mm] 
Eric A. Bergshoeff${}^1$\,,
Joaquim Gomis${}^2$ and Axel Kleinschmidt${}^{3,4}$\\[3mm]
\footnotemark[1]{\it Van Swinderen Institute,
University of Groningen,\\
Nijenborgh 4, 9747 AG Groningen, The Netherlands
}\\[1mm]
\footnotemark[2]{\it Departament de F\'isica Qu\`antica i Astrof\'isica\\
 and Institut de Ci\`encies del Cosmos (ICCUB), Universitat de Barcelona\\
Mart\'i i Franqu\`es , 08028 Barcelona, Spain}\\[1mm]
\footnotemark[3]{\it  Max-Planck-Institut f\"ur Gravitationsphysik\\
     Albert-Einstein-Institut \\
     Am M\"uhlenberg 1, 14476 Potsdam, Germany}\\[1mm]
\footnotemark[4]{\it International Solvay Institutes\\
ULB-Campus Plaine CP231, 1050 Brussels, Belgium}\\[20mm]

\begin{abstract}
\noindent
We exhibit a new method of constructing  non-Lorentzian models by applying a method we refer to as starting from a so-called seed Lagrangian. This method typically produces additional constraints in the system that can drastically alter the physical content of the model. We demonstrate our method for particles, scalars and vector fields.

\end{abstract}
\end{center}

\thispagestyle{empty}

\newpage
\setcounter{page}{1}

\setcounter{tocdepth}{2}
\tableofcontents

\section{Introduction and Summary of the Results}
\label{sec:intro}

Theories without relativistic symmetries have recently attracted a renewed interest, see for example~\cite{Grosvenor:2021hkn,Oling:2022fft,Bergshoeff:2022eog} for recent reviews.
This is motivated by their relevance to many physical situations in e.g. applications to condensed matter physics \cite{Son:2005rv,Son:2013rqa}, hydrodynamics~\cite{Ciambelli:2018xat,Poovuttikul:2019ckt,Petkou:2022bmz} and gravitational problems~\cite{Dautcourt:1996pm,Bergshoeff:2017btm,VandenBleeken:2017rij,Hansen:2019vqf,deBoer:2021jej,Campoleoni:2022ebj}. 
They have also featured prominently in non-relativistic variants of holography~\cite{Taylor:2008tg,Hartnoll:2009sz,Zaanen:2015oix}.
In this work we shall be concerned with two types of structures that arise from breaking Lorentz invariance and the interplay of these two structures.\footnote{See~\cite{Bacry:1968zf,Bacry:1986pm,Figueroa-OFarrill:2017ycu,Figueroa-OFarrill:2018ilb} for discussions and classifications of 
all possible kinematic algebras in dimensions bigger than $2+1$.}

The first instance is the Galilean limit of small velocities compared to the speed of light and where the time coordinate becomes absolute in the sense that it is not affected by the boost symmetries. The corresponding Galilei algebra is given in~\eqref{eq:Gal} below and well known to be a contraction of the relativistic Poincar\'e algebra~\cite{Inonu:1953sp}. The Galilei algebra can be centrally extended to the Bargmann algebra that also appears in many non-relativistic systems where the extension is related to a mass scale.

The second instance is the Carrollian limit where the speed of light is formally small compared to the characteristic velocities of the system
\cite{Henneaux:1979vn,Bergshoeff:2014jla,Hartong:2015xda,Bergshoeff:2017btm,Henneaux:2021yzg,deBoer:2021jej,Campoleoni:2022ebj,Hansen:2021fxi}. The associated Carroll algebra can also be obtained by a contraction of the Poincar\'e algebra~\cite{LevyLeblond:1965,gupta1966analogue} and is given in~\eqref{eq:Car}. In a Carrollian setting space becomes absolute in that it is not affected by Carrollian boosts. There is no central extension of the Carroll algebra in $D>2$ space-time dimensions. One can consider a non-central extension (by a derivation) that is dual to the Bargmann algebra. This algebraic correspondence was discussed recently in~\cite{Figueroa-OFarrill:2022pus}.

{}From the above description of these two non-Lorentzian structures it is clear that there is a formal relation between the algebras of Galilei and the Carroll (both without extensions) in $D = d+1$ space-time dimensions under the formal interchange of space and time\footnote{In~\cite{Duval:2014uoa} there is the notion of `Carroll time' $s$ that is given by $s=C ct$. In this paper, we set $c=C=1$ and therefore $s=t$ and we will always use the letter $t$ for the time coordinate.}
\begin{align}
\label{eq:xt}
\vec{x} \longleftrightarrow t\,.
\end{align}
This type of relation has been studied for example in~\cite{Barducci:2018wuj} and for $d>1$ spatial dimensions it is clearly only a formal relation.\footnote{This relation should be distinguished from the Bargmann and Carroll duality introduced in
\cite{Duval:2014uoa}.}
For $d=1$ it becomes exact and includes also the central extensions of the algebras, see for example ~\cite{Figueroa-OFarrill:2018ilb,Gomis:2020wxp}.\footnote{This type of mapping can be generalized to the case where we include corrections to the Galilei and Carroll algebras~\cite{Gomis:2022spp}. It can also be generalized  to cases with higher-dimensional foliations \cite{Barducci:2018wuj,Bergshoeff:2020xhv}.}
We discuss the case of extensions in appendix~\ref{app:massG}.

The mapping is also reflected at the level of the (repeated) action of the boost generators on the space and time coordinates. Denoting the Galilei and Carroll boosts by $\delta_{\rm G}$ and $\delta_{\rm C}$ we have that
\begin{align}
\delta_{\rm G}^2  \vec{x} &= \delta_{\rm G} ( \vec{\beta} t ) = 0 &&\text{(Galilei)}\,,\nn\\
\delta_{\rm C}^2  t &= \delta_{\rm C} ( \vec{\beta} \cdot \vec{x} ) = 0 &&\text{(Carroll)}\,,
\end{align}
where $\vec{\beta}$ labels the parameter of the (Galilei or Carroll) boost. We can also write these equations as
\begin{align}
\label{eq:nilchainx}
 \vec{x} & \longrightarrow
 t\longrightarrow0 &&\text{(Galilei)}\,,\nn\\
 t &\longrightarrow  \vec{x}\longrightarrow 0  &&\text{(Carroll)}\,.
\end{align}
under the respective boosts.
The fact that the boosts of these non-Lorentzian structures are (two-step) nilpotent is due to the contraction of the Lorentz boosts that turn an orthogonal matrix into a triangular (unipotent) one.

The central point of this paper is that one can use the formal relation between Galilei and Carroll, together with the nilpotency, to construct new non-Lorentzian systems from known ones. The connection is different from ones discussed in previous literature~\cite{Duval:2014uoa,Barducci:2018wuj,Bergshoeff:2020xhv}. 

To explain our procedure, we consider a
Carroll-invariant Lagrangian $L_{\rm C}$ that we shall refer to as a \textit{seed} for a Galilei-invariant Lagrangian.
Applying the usual  Galilei transformations $\delta_{\rm G}$ will not be an invariance of the seed Lagrangian $L_{\rm C}$ since
$\delta_{\rm G} L_{\rm C} \neq 0$.
 However, there are cases where it is possible to add terms $L_{\rm C}^\chi$ involving new \textit{Lagrange multipliers variables} $\chi$ to $L_{\rm C}$
 such that $\delta_{\rm G} (L_{\rm C} + L_{\rm C}^\chi) =0$.
The new terms added modify the dynamics, but ensure invariance of the system.
 A sufficient condition for this procedure to work is that the non-vanishing variation $\delta_{\rm G} L_{\rm C}$ of the seed Lagrangian $L_{\rm C}$ can be written as a product where one factor is invariant under the Galilei transformations and the other factor can contain terms that are not invariant.

In the non-Lorentzian context this condition is often met because of the two-step nilpotency~\eqref{eq:nilchainx} since the boost transformation of any variable is proportional to a variable with vanishing boost transformation. Therefore, we expect
to be able to find a supplementary term $L_{\rm C}^\chi$ that makes the extended system invariant under Galilei transformations. The same argument applies when taking as a starting seed a Galilei Lagrangian $L_{\rm G}$ that then is made Carroll invariant by adding a term $L_{\rm G}^\chi$ to it.

We show that this procedure of starting from a seed Lagrangian works in a variety of cases, starting from particle models, over scalar field theories to $p$-form gauge theories where also Yang--Mills is included.  Some of the models we construct in this way are new to the best of our knowledge and we give some preliminary analysis of their physical significance. It would be very interesting to extend our analysis to gravity, possibly producing new non-Lorentzian gravitational models beyond the ones already known~\cite{Hartong:2015xda,Bergshoeff:2017btm,Hansen:2021fxi,Petkou:2022bmz,Fuentealba:2022gdx,Concha:2022muu,Campoleoni:2022ebj}.

As the procedure involves Lagrange multipliers, the new dynamics has additional constraints and we therefore find non-Lorentzian models with and without constraints~\cite{Barducci:2018wuj}. In many cases we can also think of the different models as being of electric or magnetic type where this distinction is related to a dominance of time derivatives over space derivatives or vice versa. However, as will become clear from our analysis, this terminology is less unique than the distinction in terms of theories with or without constraints.

We shall also show that the models with constraints can often be related to another construction of non-Lorentzian models that naturally induces systems with Lagrangian multipliers. This construction can be called the quadratic divergence trick and was studied in~\cite{Gomis:2005pg}. The version we require applies whenever the Lagrangian has a divergence $\omega^2 X^2$ quadratic in some contraction parameter $\omega\to\infty $, with $X$ any expression in terms of the fields.
 We can control this divergence by introducing a Lagrange multiplier $\chi$ for every (component of) $X$ as follows \cite{Gomis:2005pg}:
\begin{equation}\label{quadratic}
 \omega^2 X^2\hskip 1truecm \longrightarrow \hskip 1truecm -\frac{1}{\omega^2} \chi^2 -2 \chi X\,.
 \end{equation}
Integrating out the algebraic $\chi$ from the expression on the right reproduces the original Lagrangian $\omega^2 X^2$, so that classically this is a completely equivalent description of the system before taking the limit. However, in the replaced theory we can take the limit $\omega\to \infty$ and then are left with a Lagrange multiplier term $-2\chi X$ that enforces a new constraint in the contracted theory which now has different dynamics and invariances. The relation between these two constructions will be explained in section~\ref{sec:Max}.

The structure of this paper is as follows. In section~\ref{sec:part} we study world-line models and show how the seed Lagrangian method produces new non-Lorentzian models. The same approach is then applied to scalar field theories in section~\ref{sec:KG} and to $p$-form fields in section~\ref{sec:Max}.

\section{Particle models}
\label{sec:part}

We start our discussion with the interplay of the well-known massless Galilean particle model \cite{Souriau,Batlle:2017cfa}  and the time-like Carroll particles \cite{Duval:2014uoa,Bergshoeff:2014jla} that both have the same  space-time dynamical variables~\cite{Barducci:2018wuj,Gomis:2022spp}.
Later we will see how the time-like Carroll particle can be used as a seed to construct a new massless Galilean particle. The construction is possible due to the introduction of Lagrange multiplier variables that restrict the dynamics of the new massless Galilean particle. As we will see an analogous construction for a new timelike Carroll particle can be done starting with the
ordinary massless Galilei particle.
An analysis of the massive Galilei particle, relying on the centrally extended Galilei algebra, can be found in appendix~\ref{app:massG}.

We work in flat $D$-dimensional space-time with signature $(-++\ldots+)$ and indices $a=1,\ldots,d$ label the (flat) spatial directions, where $D=1+d$.
The Planck constant, the velocity of light $c$ and
its Carrollian analogue $C$~\cite{LevyLeblond:1965,Duval:2014uoa} will be set to one, $\hbar=c=C=1$ in this work.

\subsection{Magnetic massless Galilei particle }
\label{sec:MGP}

We start by considering the spinless massless Galilei particle  that depends on the ordinary phase space variables. It  carries so-called `colour'~\cite{Souriau} that we denote by $m$.
The canonical Lagrangian is (see also \cite{Batlle:2017cfa})
\begin{align}
\label{eq:MLG0}
L_{\rm mG} &= - E \dot{t} + \vec{p} \cdot \dot{\vec{x}}
 -\frac{e}{2} \left( \vec{p}^{\, 2} - m^2 \right)
 \,,
 \end{align}
 the dot derivative here is with respect to a dimensionless world-line parameter $\tau$.
 The mass-shell constraint only depends on the spatial momentum, this is the reason we call the particle a \textit{magnetic} massless Galilei particle.
The Lagrangian~\eqref{eq:MLG0} is invariant under the Galilei transformations
\begin{subequations}
\label{eq:TRG0}
\begin{align}
\delta_{\rm G} x^a &= \epsilon^a -\omega^{ab}  x_{ b} + \beta^a {t}\,, &
\delta_{\rm G} t&= \eta \,,\\
\delta_{\rm G} p^a &= -\omega^{ab} p_{ b} \,, &
\delta_{\rm G} E &=   \beta^a p_{a}\,,
\end{align}
\end{subequations}
along with $\delta_{\rm G} e=0$. The transformation of the momentum variables $(E,\vec{p})$ follows from invariance of the symplectic structure terms of \eqref{eq:MLG0}. The parameters $(\epsilon^a, \, \eta,\, \beta^a\,, \omega^{[ab]})$ are constant and parametrise spatial translations, time translations, Galilei boosts and spatial rotations, respectively.
In analogy with~\eqref{eq:nilchainx} we see that we have a two-step nilpotency
\begin{align}
\label{eq:nilG}
E \to \vec{p} \to 0
\end{align}
under Galilei boosts.
Note in particular that $\delta_{\rm G} \vec{p}=0$ under boosts and that the momenta are invariant under the translations $\epsilon^a$ and $\eta$.

 The algebra of symmetries of the magnetic Galilei particle is the Galilei algebra without central extension
\begin{align}
\label{eq:Gal}
[ J_{ab} , J_{cd} ] &= \delta_{bc} J_{ad} - \delta_{ac} J_{bd} - \delta_{bd} J_{ac} + \delta_{ad} J_{bc} \,,\nn\\
[ J_{ab} , B_c ] &= \delta_{bc} B_a - \delta_{ac} B_b\,,\nn\\
[ J_{ab} , T_c] &=  \delta_{bc} T_a - \delta_{ac} T_b\,,\nn\\
[ J_{ab} , H ] &=0\,,\nn\\
[ B_a , B_b ] &= 0 \,,\nn\\
[ B_a , T_b ] &= 0\,,\nn\\
[ B_a, H ] &= - T_a\,,\nn\\
[ T_a, T_b ] &= [T_a, H] =0\,,
\end{align}
where $J_{ab}$ (anti-symmetric parameter $\omega_{ab}$) are the $\mathfrak{so}(d)$  spatial rotations, $B_a$ (parameter $\beta^a$) the commuting Galilei boosts, $T_a$ (parameter $\epsilon^a$) the spatial translations and the Hamiltonian $H$ (parameter $\eta$) corresponds to time translations.

Even though it is standard, we briefly present the canonical analysis for the Lagrangian~\eqref{eq:MLG0}.
Denoting all conjugate momenta with a letter $\pi$, we have the primary constraints\footnote{We follow the convention that $\pi_t = - \frac{\partial L}{\partial \dot{t}}$ and $\pi_E = - \frac{\partial L}{\partial \dot{E}}$, while the other conjugate momenta are defined with a plus sign. These definitions are  due the symplectic part of the canonical actions.}
\begin{align}
\label{eq:prim1}
\pi_E = \pi_e = \pi^a_p =0\,,\quad \pi_t = E\,, \quad \pi_x^a = p^a
\end{align}
The final Dirac Hamiltonian is 
\begin{align}
H_D=\frac{e}{2} \left( \vec{p}^{\, 2} - m^2 \right)
 +\pi_e \lambda(\tau),
 \end{align}
where $\lambda(\tau)$ is an arbitrary function of $\tau$. Stability of the primary constraints
 implies the secondary constraint $\vec{p}^{\, 2} - m^2 =0$ from the evolution of $\pi_e=0$. The other primary constraints in~\eqref{eq:prim1} form second-class pairs and therefore do not appear in the Dirac Hamiltonian.
 Working in $d$ spatial dimensions there are then $4+2d$ constraints in total among which there are $2$ first-class constraints corresponding to world-line reparametrisation invariance. Therefore the total count of degrees of freedom in phase space is
 $2\times (3+2d) - 2\times 2 - (2+2d) = 2\times d$. This number is composed out of the $2\times(d-1)$ components of the $(\vec{x},\vec{p})$ sector (since $\vec{p}^{\, 2}=m^2$) and the $2\times 1$ degrees of freedom of the pair $(t,E)$ and these two sectors decouple.

An important comment on the interpretation of the $2\times(d-1)$ degrees of freedom in the $(\vec{x},\vec{p})$ sector is in order. Since the constraint $\vec{p}^{\, 2} = m^2$ is in Euclidean space, making it second-class is inconsistent with fixing the reparametrisation to $\tau=t$. Rather one has to consider a `Euclidean evolution' $\tau= x^1$ (say) and the reduced $d-1$ degrees of freedom can be considered as transverse to this choice of Euclidean direction, for an analogous situation for a non-relativistic non-vibrating string  
see~\cite{Batlle:2016iel,Gomis:2016zur}.

We note that we could replace the mass-shell constraint $\vec{p}^{\, 2}-m^2=0$ by an arbitrary potential $V(\vec{p}^{\,2})$ that would lead to the same counting of  degrees of freedom (as long as the derivative of the potential is non-zero at the origin). Since $\vec{p}$ is Galilei boost invariant, this modification also preserves non-relativistic Galilei invariance.

\subsection{Electric massive Carroll particle}
\label{sec:ECP}

The canonical Lagrangian for a time-like massive Carroll particle \cite{Bergshoeff:2014jla,Duval:2014uoa} is
\begin{align}
\label{eq:EC0}
L_{\rm eC} &= - E \dot{t} + \vec{p} \cdot \dot{\vec{x}}
-\frac{e}{2} \left( -E^2+ m^2 \right)\,.
\end{align}
where $t$ is the Carrollian time~\cite{LevyLeblond:1965,Duval:2014uoa} and we recall that we set $c=C=1$. We call~\eqref{eq:EC0} the electric Carroll particle since the mass-shell
constraint only depends on the energy.
The Lagrangian~\eqref{eq:EC0} is invariant under the Carroll transformations
\begin{subequations}
\label{eq:TRC0}
\begin{align}
\delta_{\rm C} x^a &= \epsilon^a  - \omega^{ab} x_{b} \,, &
\delta_{\rm C} p^a &= - \omega^{ab} p_{ b} + \beta^a \,{E} \,,\\
\delta_{\rm C}\, t &= \eta + \vec\beta\cdot \vec{x}\,, &
\delta_{\rm C} {E} &=0
\,.
\end{align}
\end{subequations}
The einbein is invariant $\delta_{\rm C} e=0$.
In analogy with~\eqref{eq:nilchainx} we have the two-step nilpotency relation
\begin{align}
\label{eq:nilC}
\vec{p} \to E \to 0
\end{align}
under Carroll boosts.

The algebra of these transformations is the Carroll algebra
\begin{align}
\label{eq:Car}
[ J_{ab} , J_{cd} ] &= \delta_{bc} J_{ad} - \delta_{ac} J_{bd} - \delta_{bd} J_{ac} + \delta_{ad} J_{bc} \,,\nn\\
[ J_{ab} , K_c ] &= \delta_{bc} K_a - \delta_{ac} K_b\,,\nn\\
[ J_{ab} , T_c ] &=  \delta_{bc} T_a - \delta_{ac} T_b\,,\nn\\
[ J_{ab} , H ] &=0\,,\nn\\
[ K_a , K_b ] &= 0 \,,\nn\\
[ K_a , T_b ] &= -  \delta_{ab} H\,,\nn\\
[ K_a, H ] &= 0\,,\nn\\
[ T_a, T_b ] &= [T_a, H] =0\,.
\end{align}
The Carrollian boosts are called $K_a$ in order to distinguish them from the Galilean ones.

The Carroll Lagrangian~\eqref{eq:EC0} and Carroll transformations~\eqref{eq:TRC0} can be obtained from the corresponding ones of the magnetic massless Galilei case (see~\eqref{eq:MLG0}, \eqref{eq:TRG0}) under the map~\cite{Barducci:2018wuj,Gomis:2022spp}
\begin{align}
\label{eq:map}
t\longleftrightarrow \vec x\,,
\quad\quad \vec p
\longleftrightarrow -E\,.
\end{align}

Besides the global transformations~\eqref{eq:TRC0}, the Lagrangian also enjoys a gauge-invariance associated with world-line reparametrisations. In order to exhibit this and to illustrate the physical content, we perform a canonical analysis of~\eqref{eq:EC0}. We assume $e\neq 0$ and $E\neq 0$ to work at a generic point in phase space.

Denoting the canonical momenta of the variables by $\pi$, we have the primary constraints
\begin{align}
\pi_E = 0 \,, \quad
\pi_t - E = 0\,,\quad
\pi_p^a = 0 \,,\quad
\pi_x^a - p^a =0\,,\quad
\pi_e =0\,.
\end{align}
The final Dirac Hamiltonian is
\begin{align}
H_D = \frac{e}{2}\left( -E^2 +m^2 \right) + \lambda_e \pi_e
\end{align}
and the stability of the primary constraints leads to the secondary constraint
\begin{align}
 -E^2 +m^2 = 0\,.
\end{align}
There are no further constraints. The total number of constraints is therefore $4+2d$ if one also takes the vector indices in $d$ space dimensions into account. Among these, two are first-class and the remaining $2+2d$ constraints are second-class. This leads to the expected total number of $2\times(3+2d)-2\times 2 - (2+2d)= 2 \times d $ physical degrees of freedom in phase space. These are completely due to the variables $(\vec{x},\vec{p})$. The equations of motion of~\eqref{eq:EC0} indeed force $\vec{x}$ and $\vec{p}$ to be constants whose arbitrary values parametrise the physical phase space. The constancy of $\vec{x}$ reflects the well-known fact that a time-like Carroll particle does not move.

Taking into account also the gauge transformation of the Lagrange multiplier, we arrive the following gauge transformations of the variables appearing in~\eqref{eq:EC0}:
\begin{align}
\delta_\lambda e &= \dot{\lambda} \,,&
\delta_\lambda t &= \lambda E
\end{align}
while the other variables in~\eqref{eq:EC0} are gauge-invariant.
The generator of the gauge transformation is
\begin{align}
G=\dot\lambda\,\pi_e+\frac{\lambda}{2}
(-E^2 +m^2)\,.
\end{align}

Similar to the case of the magnetic massless Galilei particle, we could also allow for more general dispersion relations of the form $V(E)-m^2=0$. This does not influence the counting of degrees of freedom as long as $V(E)$ is not constant. Moreover, it preserves Carroll invariance since $\delta_{\rm C} E =0$.

\subsection{Electric massless Galilei particle}
\label{sec:EGP}

We now come to the application of the strategy outlined in the introduction that generates new non-relativistic Lagrangians from seed Lagrangians.
For a two-step nilpotency of the form~\eqref{eq:nilG} and canonical Lagrangians of the form
\begin{align}
L_{\rm C} = - E\dot{t} + \vec{p}\cdot \dot{\vec{x}} - H(e,E,p)
\end{align}
the strategy is guaranteed to work in general since
\begin{align}
\delta_{\rm G} L_{\rm C} = - \frac{\partial H}{\partial E} \delta_{\rm G} E = -  \frac{\partial H}{\partial E} \beta^a p_a = - \delta_{\rm G} ( \chi^a p_a )
\end{align}
for $\delta_{\rm G} \chi^a = \beta^a   \frac{\partial H}{\partial E}$. Therefore $L_{\rm G} = L_{\rm C} + \chi^a p_a$ is Galilei invariant. A similar reasoning applies when starting from a Galilei invariant $L_{\rm G}$ of a similar form due to the two-step nilpotency~\eqref{eq:nilC}.

Our first example to illustrate this is to take as the seed Lagrangian the electric massive Carroll  particle~\eqref{eq:EC0} with constraint $E^2- m^2 =0$.
Clearly this mass-shell constraint is not invariant under the Galilei transformations~\eqref{eq:TRG0} since
\begin{align}
\delta_{\rm G} (E^2- m^2 )=2E\vec\beta\cdot{\vec{p}}\,.
\end{align}
However, it is proportional to $\vec{p}$ for which we know that $\delta_{\rm G}{\vec{p}}=0$, see~\eqref{eq:nilG}. We can therefore construct a new invariant mass-shell constraint by adding a Lagrange multiplier $\vec\chi$ term to the constraint via
\begin{align}
\label{eq:GnC}
E^2- m^2 -2\vec\chi\cdot {\vec{p}}=0\,.
\end{align}
If the transformation of the new variable $\vec{\chi}$ is given by
\begin{align}
\label{eq:chiG0}
\delta_{\rm G} \vec{\chi}= E\vec\beta\,,
\end{align}
the new constraint~\eqref{eq:GnC} is Galilei boost invariant.
Note that in this case in order to have the Galilei invariance
of the canonical Lagrangian we need to introduce a Lagrange multiplier variable $\vec\chi$ that restricts the dynamics to $\vec p=0$.
This restriction of the dynamics is a general feature of the procedure based on the nilpotency of the transformations.
Note moreover that the new variable $\vec{\chi}$ extends the two-step nilpotent chain to a three-step chain:
\begin{align}
    \vec\chi \to E \to \vec{p} \to 0\,.
\end{align}

With this new Lagrange multiplier we can write the action of the electric massless Galilei particle as
\begin{align}
\label{eq:EG0}
L_{\rm eG} &= - E\dot{t} + \vec{p} \cdot \dot{\vec{x}}
 -\frac{e}{2} \left(E^2 - m^2 -2\vec\chi\cdot \vec {p}\right)
 \,.
\end{align}

In order to elucidate the dynamical content of the Lagrangian~\eqref{eq:EG0} we again perform a canonical analysis. For simplicity, we make the field redefinition $\tilde\chi^a \equiv e \chi^a$ in order to disentangle the two Lagrange multipliers.
The canonical momenta dual to the variables $(E,t,p^a, x^a, e, \tilde{\chi}^a)$ are all constrained by primary constraints that read explicitly
\begin{align}
\pi_E = 0 \,, \quad
\pi_t - E =0 \,,\quad
\pi_p^a =0\,, \quad
\pi_x^a -p^a =0\,,\quad
\pi_e =0\,,\quad
\pi_{\tilde{\chi}}^a =0\,.
\end{align}
There are two secondary constraint given by
\begin{align}
E^2-m^2  = 0\,,\quad
 p^a =0\,.
\end{align}
There are no further constraints and we work at a generic point in phase space, meaning that we assume that $e\neq 0$ and $\vec{p}\neq 0$. Among the total of $4+4d$ constraints there are $2+2d$ first-class and $2+2d$ second-class constraints. This leads to a total number of $2\times(3+3d)-2\times (2+2d) -(2+2d)=0$ degrees of freedom. Thus, there are no propagating dynamics contained in the Lagrangian~\eqref{eq:EG0}.
This result is not surprising since the Lagrange multipliers $e$ and $\vec{\chi}$ force all variables to constant values and there are no non-trivial solutions of the equations of motion then. Yet another way of arriving at this conclusion is by integrating out the fields that appear algebraically by using their equations of motion. After integrating out $(E,\chi^a, p^a, e)$. the Lagrangian reduces to a total derivative, also showing that there are no dynamics contained in it.

For completeness, we also record the gauge transformations of~\eqref{eq:EG0}:
\begin{align}
\delta_{\lambda, \vec{\sigma}} e = \dot{\lambda}\,,\quad
\delta_{\lambda, \vec{\sigma}} t= - \lambda E \,,\quad
\delta_{\lambda, \vec{\sigma}} \tilde\chi^a = \dot{\sigma}^a\,,\quad
\delta_{\lambda, \vec{\sigma}} x^a = -\sigma^a\,,\quad
\delta_{\lambda, \vec{\sigma}} p^a = 0\,,\quad
\delta_{\lambda, \vec{\sigma}} E = 0\,.
\end{align}
Here, $\lambda$ and $\sigma^a$ are arbitrary functions of the world-line parameter.

We note that, since the equation of motion for $\vec{\chi}$ enforces $\vec{p}=0$ on-shell, the transformation of $E$ under a Galilei boost is zero on-shell, see~\eqref{eq:TRG0}.  This means that we can define a modified Galilei boost by adding a trivial transformation, following the general pattern $\tilde{\delta} \varphi_I =   \delta \varphi_I + a_{IJ} \frac{\delta L}{\delta \varphi_J}$ for any anti-symmetric $a_{IJ}$, where $\varphi_I$ ranges over all fields in the action.
In the present case, we can arrange this to arrive at
\begin{align}
\tilde{\delta}_{\rm G} E = 0 \,, \quad \tilde{\delta}_{\rm G} \vec{\chi} = -\vec{\beta} \frac{\dot{t}}{e}\,.
\end{align}
All other transformations in~\eqref{eq:TRG0} remain unchanged.
Since now $E$ is invariant under this modified Galilei boost, there is no longer any reason to use the free non-relativistic dispersion relation $E^2-m^2$ and we could substitute $E$ by an arbitrary function $V(E)$ while maintaining invariance under Galilei boosts. This again does not modify the invariances or the counting of degrees of freedom.

More generally, the Lagrangian~\eqref{eq:EG0} is invariant under the transformations
\begin{align}
\delta_{\rm G} \chi^a = E\beta^a - \omega^{ab} \chi_b - \omega_{(1)}^{ab} p_{b}\,,
\end{align}
where $\omega^{ab}$ simply represents the usual spatial rotations while $\omega_{(1)}^{ab}=\omega_{(1)}^{[ab]}$ is a seemingly new transformation that does not act on $(E, p^a)$. It is trivially a symmetry since $L$ varies into $\omega_{(1)}^{ab} p_{a} p_{b} =0$ under it.
Since $p^a=e^{-1} \frac{\delta L}{\delta \chi_a} $ is proportional to an anti-symmetric combination of the equations of motion, the symmetry is actually a trivial, or zilch, symmetry of the system. It also arises in the commutator of two boost transformations on $\chi$:
\begin{align}
\label{eq:com1}
\left[ \delta_{\beta_1} , \delta_{\beta_2}\right] \chi^a &=
(\beta_2^a\beta_1^b-\beta_1^a\beta_2^b) p_{b} = \delta_{\sigma_{(1)}} \chi^a
\end{align}
with $\sigma_{(1)}^{ab} = 2\beta_1^{[a}\beta_2^{b]}$.
We can think of~\eqref{eq:com1} as an example of an open algebra.

\subsection{Magnetic massive Carroll particle}
\label{sec:MCP}

Now we will construct a Carroll invariant particle with Lagrange multiplier starting from
the magnetic Galilei invariant Lagrangian~\eqref{eq:MLG0}
as a seed.
The mass-shell constraint is
$\vec{p}^{\, 2} - m^2=0$.
It is not invariant under the Carroll transformations (\ref{eq:TRC0})
\begin{align}
\delta_{\rm C} (\vec p^{\, 2}- m^2 )=
2\vec p\cdot\vec\beta\, E\,,
\end{align}
but is proportional to ${{E}}$ which is invariant under Carroll transformation, see~\eqref{eq:nilC}. Therefore we can again add a Lagrange multiplier term to obtain a magnetic massive Carroll-invariant action
\begin{align}
\label{eq:MC0}
L_{\rm mC} &= - E \dot{t} + \vec{p} \cdot \dot{\vec{x}} -\frac{e}{2} \left( \vec p^{\, 2}- m^2 -2\chi\cdot {E}\right)\,.
\end{align}
The Carroll boost transformation of $\chi$ is given by
\begin{align}
\delta_{\rm C} \chi = \vec p\cdot\vec\beta\,,
\end{align}
again giving rise to a three-step nilpotency
\begin{align}
\chi\to\vec{p}\to E\to 0\,.
\end{align}

We again perform a canonical analysis after redefining the Lagrange multiplier as $\tilde\chi = e \chi$ and working at a generic point in phase space, in particular $e\neq 0$.  All variables $(E, t, p^a, x^a, e, \tilde{\chi})$ give rise to primary constraints
\begin{align}
\pi_E=0\,,\quad
\pi_t-E=0\,,\quad
\pi_p^a=0\,,\quad
\pi_x^a -p^a=0\,,\quad
\pi_e=0\,,\quad
\pi_{\tilde\chi}=0\,.\quad
\end{align}
Their stability entails the secondary constraints
\begin{align}
\vec{p}^{\, 2}-m^2 =0 \,,\quad
 E = 0\,.
\end{align}
Among the total of $6+2d$ constraints there are $4$ first-class constraints and $2+2d$ second-class constraints. This leads to a total of $2\times(4+2d)-2\times 4 -(2+2d) = 2\times (d-1)$ propagating degrees of freedom as expected from the Carroll tachyon~\cite{deBoer:2021jej,Gomis:2022spp}.
This can be again be interpreted in terms of a Euclidean evolution, similar to the magnetic massless Galilei case.
By contrast, the magnetic massless Galilei particle had $2\times d$ degrees of freedom since the energy $E$ could take  any constant value there, whereas here it is constrained to $E=0$.

We also record the gauge-invariances of~\eqref{eq:MC0}
\begin{align}
\delta_{\lambda,\sigma} e =  \dot\lambda\,,\quad
\delta_{\lambda,\sigma} t =  \lambda\,,\quad
\delta_{\lambda,\sigma} \tilde\chi =  \dot\sigma\,,\quad
\delta_{\lambda,\sigma} x^a = \sigma p^a\,,\quad
\delta_{\lambda,\sigma} p^a =  0\,,\quad
\delta_{\lambda,\sigma} E =  0\,.
\end{align}

Similarly to the electric Galilei case, we could introduce modified Carroll boosts $\tilde{\delta}_{\rm C}$ by adding trivial transformations such that
\begin{align}
\tilde{\delta}_{\rm C} \vec{p} = 0 \,,\quad\quad
\tilde{\delta}_{\rm C} \chi = - \frac{\vec{\beta}\cdot \vec{x}}{e}\,.
\end{align}
Moreover, there is the possibility of using a modified dispersion relation by replacing $\vec{p}^{\, 2}$ in the mass-shell constraint by an arbitrary function of $\vec{p}^{\, 2}$.

The commutator of two Carroll boosts on $\chi$ is
\begin{align}
\left[ \delta_{\beta_1} , \delta_{\beta_2} \right] \chi = 0\,.
\end{align}
There is no new transformation occurring in this case, unlike~\eqref{eq:com1} where a trivial transformation arose. The symmetry algebra of the magnetic massive Carroll particle is therefore the usual Carroll algebra~\eqref{eq:Car} and is closed, just like in the electric case.


\section{From the world-line to field theory}
\label{sec:KG}

In order to construct scalar field theories from the above particle Lagrangians, in particular from the mass-shell constraints,
we realise the momenta as differential operators on the scalar field $\phi$.
In particular, we will think of $p_a$ and $E$ as derivatives $-i \partial_a \phi$ and $ i \dot{\phi}\equiv i\partial_t \phi$.
The field theories without Lagrange multipliers arise from enforcing the
mass-shell constraints by sandwiching with the real scalar field.

\subsection{Magnetic Galilei scalar field}

The scalar field theory corresponding to~\eqref{eq:MLG0} is obtained  from the constraint $\vec{p}^{\, 2} - m^2=0$. In fact, following the Dirac procedure
the constraint implies the wave equation for the scalar field $\phi(t, \vec x)$
\begin{align}
\label{eq:mgeom}
\left(\partial_a \partial^a + m^2\right)
\phi(t, \vec x)=0\,.
\end{align}
This wave equation can be derived
 from the Lagrangian
\begin{align}\label{maggal}
\cL_{\rm mG} = \frac12 \phi(t,\vec x) \left( \partial_a \partial^a + m^2 \right) \phi(t,\vec x) = - \frac12 (\partial_a \phi(t,\vec x))^2 + \frac12 m^2 \phi(t,\vec x)^2\,.
\end{align}
This Lagrangian can also be obtained from the non-relativistic limit of a tachyonic relativistic Klein--Gordon field \cite{Bergshoeff:2022eog}.
The Galilei boost $\delta_{\rm G} E = \vec{\beta} \cdot \vec{p}$ from~\eqref{eq:TRG0} translates into the following transformation of the field:
\begin{align}
\label{eq:phiG0}
\delta_{\rm G} \phi = t\, \beta^a\partial_a  \phi \,,
\end{align}
which is the corresponding transport term associated to the Galilean boosts in configuration space.
For completeness, we also list the transformation of the field under the remaining elements of the Galilei algebra, each parametrised by their own parameters,
\begin{subequations}
\label{GKG}
\begin{align}
\delta_{B}\phi& = t \,\beta^a \partial_a \phi   \,, &
 \delta_{T}  \phi&= \epsilon^a \partial_a \phi \,,\\
 \delta_{H}   \phi_{(0)}  & = \eta \partial_t \phi\,, &
 \delta_{J}  \phi_{(0)} & =\omega^{ab}x_a\partial_b \phi\,.
\end{align}
\end{subequations}

Let us remark on the fate of the nilpotency~\eqref{eq:nilchainx} of the Galilei transformations in field theory. Clearly, the behaviour~\eqref{eq:phiG0} under boosts is not two-step nilpotent. But since this is solely the transport term of the nilpotent Galilei boost in configuration space, this is the correct field theory analogue. This fact will be crucial for finding another Galilei-invariant theory below.

We also note that, since the variable $t$ is a spectator variable in the Lagrangian~\eqref{maggal}, the proper evolution equation is to be thought of as Euclidean, similar to the particle case in section~\ref{sec:MGP}.

\subsection{Electric Carroll scalar field}

The scalar Carroll field theory corresponding to~\eqref{eq:EC0} is obtained  from the constraint $E^{\, 2} - m^2=0$.
The wave equation for the scalar field $\phi(t, \vec x)$ is
\begin{align}
\left(\partial_t ^{\, 2}+ m^2\right)
\phi(t, \vec x)=0,
\end{align}

The Lagrangian of scalar field theory corresponding to the
massive Carroll particle Lagrangian~\eqref{eq:EC0} is
\begin{align}\label{carrollKG}
\mathcal{L}_{\rm eC} = \frac12 \phi \left[ -\partial_t^2-m^2 \right] \phi =  \frac12 (\partial_t\phi)^2 - \frac12 m^2 \phi^2
\end{align}
 so that this dynamics is unrestricted.
 This Lagrangian can also be obtained from a non-relativistic limit of a relativistic massive Klein--Gordon field \cite{Bergshoeff:2022eog}.

 The Lagrangian is quasi-invariant under the Carroll transformations
\begin{subequations}
\label{eleccarrolltrans}
 \begin{align}
\delta_{K}\phi& = \vec x \,\cdot\vec\beta \partial_t \phi   \,, &
 \delta_{T}  \phi&= \epsilon^a \partial_a \phi \,,\\
 \delta_{H}   \phi  & = \eta \partial_t \,, &
 \delta_{J}  \phi & =\omega^{ab}x_a\partial_b \phi\,.
\end{align}
\end{subequations}

A canonical analysis here gives two physical degrees of freedom in phase space which is reasonable in view of section~\ref{sec:ECP}.

\subsection{Electric Galilei scalar field}

We now apply the strategy outlined in the introduction to the Galilei and Carroll field theories above.
 The seed Lagrangian for obtaining another Galilei theory is the Carroll scalar Lagrangian
\eqref{carrollKG} which is not invariant under the Galilei transformations \eqref{GKG}. In order to get Galilei invariance as for the particle case we add the field theory analogue of the term $\vec\chi\cdot\vec p$ to the seed Lagrangian. We have
\begin{align}
\label{eq:EGS}
\cL_{\rm eG} = \frac12 \phi \left(- \partial_t^2 - m^2 \right) \phi - \chi^a \partial_a \phi =  \frac12 \dot{\phi}^2 - \frac12 m^2 \phi^2 - \chi^a \partial_a \phi \,.
\end{align}
This is (quasi-)invariant under~\eqref{eq:phiG0} and
\begin{align}
\delta_{\rm G} \chi^a &= t\, \beta^b \partial_b \chi^a + \beta^a \dot{\phi}\,.
\end{align}
The transformation of $\chi^a$ can be understood as follows. The first term is the `translation term' that comes from the transformation of the argument of the field just as in~\eqref{eq:phiG0} while the second term can be understood from~\eqref{eq:chiG0}.
This is in agreement with the corresponding construction in the particle model and the reinterpretation of the nilpotency of the boost transformation. More explicitly, ignoring the transport term, we have
\begin{align}
   \chi^a \to \dot{\phi} \to \partial_a \phi \to 0\,.
\end{align}

Instead of applying the canonical formalism to study the number of degrees of freedom (per spatial point) of the Lagrangian~\eqref{eq:EGS}, we look at the solutions of the equations of motion
\begin{align}
\ddot{\phi} + m^2 \phi - \partial_a \chi^a = 0\,,\quad
\partial_a \phi =0\,.
\end{align}
The second of these implies that $\phi$ is only a function of time $t$ which then by the first implies that $\partial_a \chi^a$ is also only a function of time. The most general solution can then be written as
\begin{align}
\label{eq:EGSsol}
\phi(t,\vec{x}) = \phi(t) \,,\quad
\chi^a (t,\vec{x})  = \frac{1}{3} x^a\left(\ddot{\phi}(t) +m^2\phi(t)\right) + \alpha^a(t) + \varepsilon^{abc} \partial_{b} \gamma_c(t,\vec{x})\,,
\end{align}
where $\phi(t)$ and $\alpha^a(t)$ are arbitrary functions of time while $\gamma_a(t,\vec{x})$ is an arbitrary function of space and time and we have written this term for $D=3+1$ for simplicity.

The Lagrangian~\eqref{eq:EGS} has a gauge invariance under the local transformations
\begin{align}
\delta_\lambda \phi = \partial_a \lambda^a\,,\quad
\delta_\lambda \chi^a =  \ddot{\lambda}^a+ m^2 \lambda^a\,,
\end{align}
where the gauge parameter has to satisfy $\partial_a \partial_b \lambda^b(t,\vec{x})=0$. This in turn means that
\begin{align}
\lambda^a (t,\vec{x})  = \frac{1}{3} x^a f(t) + g^a(t) + \varepsilon^{abc} \partial_b h_c(t,\vec{x})\,,
\end{align}
which implies that the solution~\eqref{eq:EGSsol} is gauge-equivalent to the trivial solution. Therefore, there are no local bulk physical degrees of freedom contained in the field theory described by~\eqref{eq:EGS}. This is in agreement with the particle model analysed in section~\ref{sec:EGP}. It would be interesting to investigate the possible boundary degrees of freedom that can arise by imposing appropriate boundary conditions.

\subsection{Magnetic Carroll scalar field}

In this case the seed Lagrangian is the Galilei invariant scalar theory \eqref{maggal}, which  is not invariant under the Carroll transformation \eqref{eleccarrolltrans}. In order to get Carroll invariance we add the
field theory
analogue of the term $\chi\,E$ in the particle case. We have
\begin{align}\label{case1bflat}
\mathcal{L}_{\text{mC}}=   - \frac12\,  (\partial_{a}\phi)^2   +\frac12 m^2 \phi^2-\chi\partial_{t}\phi\,.
\end{align}
Note that we have a restriction on the dynamics due to presence of the variable $\chi$.

The boost transformations are
\begin{align}
\delta _{\rm C}\phi = \vec\beta \cdot\vec x \, \partial_t \phi  \,, \quad\quad
\delta_{\rm C} \chi = \vec\beta \cdot\vec x \,\partial_t \chi- \beta^i \partial_i \phi\,.
\end{align}
and leave the Lagrangian quasi-invariant.
The first terms in the transformations of $\phi$
and $\chi$ correspond to the transport terms
due to the Carroll transformation. The second term of the transformation of $\chi$ is the field theory implementation of the particle transformation of the variable $\chi$, \,$\delta_{\rm C} \chi = \vec p\cdot\vec\beta$.

Note that in this case the Lagrange multiplier can be understood as the momentum
of $\phi$, $\chi=-\pi_\phi$, and the action
\eqref{case1bflat} can be written as
\begin{align}
\mathcal{L}_{\text{mC}} =\pi_\phi \partial_t \phi -\frac12 \left((\partial_{a}\phi)^2-m^2\phi^2\right)
\end{align}
which for $m=0$ is an agreement with the magnetic Carroll Hamiltonian action of \cite{Henneaux:2021yzg}.

Let us also consider the equations of motion associated with the field theory~\eqref{case1bflat}:
\begin{align}
\partial_a \partial^a \phi + m^2 \phi+ \dot \chi = 0 \,,\quad
\dot{\phi} = 0\,.
\end{align}
The last constraint signifies that $\phi(t,\vec{x})$ only depends on $\vec{x}$ and therefore $\dot{\chi}$ also only depends on $\vec{x}$. The most general solution can then be written as
\begin{align}
\phi(t,\vec{x}) = \phi(\vec{x})\,,\quad
\chi(t,\vec{x}) =- t \left(\partial_a \partial^a \phi(\vec{x}) +m^2 \phi(\vec{x}) \right)+ a(\vec{x})
\end{align}
for arbitrary $\phi(\vec{x})$ and $a(\vec{x})$ of the spatial position.

The action~\eqref{case1bflat} has no gauge invariance and therefore there are two independent quantities per spatial point. The theory therefore describes propagating degrees of freedom just as the particle model discussed in section~\ref{sec:MCP}.


\section{\texorpdfstring{Vector fields and $p$-forms}{Vector fields and p-forms}}
\label{sec:Max}

In this section we extend the discussion from the scalar field theories in the previous section to vector field theories  and, even more, to non-linear Yang-Mills theories and $p$-form field theories. We will take a slightly different approach here where we consider Maxwell theories in the presence of gravity.

Our starting point  is the following Lagrangian describing Maxwell's theory coupled to general relativity:
\begin{equation}
  \label{eq:realvecrel}
  E^{-1} \mathcal{L}_{\rm rel} = - \frac14\, E^\mu{}_{ A} E^{\rho  A}  E^\nu{}_{ B} E^{\sigma B} F_{\mu\nu} F_{\rho \sigma} \,.
\end{equation}
Here $E_\mu{}^A$ is the Vierbein field, $E = \det E_\mu{}^A$ and $F_{\mu\nu}= \partial_\mu A_\nu - \partial_\nu A_\mu$ is the usual Maxwell field strength. This Lagrangian corresponds  to the following equations of motion and Bianchi identities:
\begin{equation}\label{e.o.m.}
D^B F_{BA} = 0\,,\hskip 2truecm D_{[A} F_{BC]}=0\,,
\end{equation}
where $D_A= E_A{}^\mu D_\mu$ and $D_\mu$ represents a Lorentz-covariant derivative with $D_\mu F_{AB} = \partial_\mu F_{AB} + 2 \Omega_{\mu[A}{}^C F_{B]C}$.
The action corresponding to the Lagrangian \eqref{eq:realvecrel} and the equations of motion \eqref{e.o.m.}  are invariant under the following general coordinate transformations, local Lorentz rotations and $U(1)$ gauge transformations with parameters $\xi^\mu, \Lambda^A{}_B$ and $\Lambda$, repectively:
\begin{subequations}
\begin{eqnarray}
\delta A_\mu &=& \xi^\lambda \partial_\lambda A_\mu + \partial_\mu \xi^\lambda A_\lambda + \partial_\mu \Lambda\,,\\[.1truecm]
\delta E_\mu{}^A &=& \xi^\lambda\partial_\lambda E_\mu{}^A + \partial_\mu\xi^\lambda E_\lambda{}^A + \Lambda^A{}_B E_\mu{}^B\,.
\end{eqnarray}
\end{subequations}
which imply the transformation rules
\begin{subequations}
\begin{eqnarray}
\delta F_{\mu\nu} &=& \xi^\lambda\partial_\lambda F_{\mu\nu} + 2\partial_{[\mu}\xi^\lambda F_{\lambda \nu]}\,,\\[.1truecm]
\delta E &=& \partial_\lambda (E\xi^\lambda)\,.
\end{eqnarray}
\end{subequations}

To define a non-Lorentzian limit we will first redefine the gravitational fields and the Maxwell gauge field  thereby introducing a (dimensionless) contraction parameter $\omega$ and next take the limit by taking $\omega \to \infty$.
Making use of these redefinitions  we first consider the vector field analogue of the magnetic Galilei and electric Carroll scalar field theories discussed in the previous section. They lead to non-Lorentzian theories without a Lagrange multiplier and, therefore, without constraints. Next, we will consider the analogues of the electric Galilei and magnetic Carroll scalar field theories that do contain a Lagrange multiplier. We will show that the results obtained by taking a limit using the  quadratic divergence trick \eqref{quadratic} are the same as the ones that can be obtained by applying the seed Lagrangian method.

\subsection{Magnetic Galilean Maxwell}

To define the Galilean limit  and derive the resulting symmetries, we redefine the Vierbein field $E_\mu{}^A$ and the Lorentz parameters $\Lambda^{AB}$ as follows\cite{Bergshoeff:2017btm}:
\begin{subequations}
\label{Galileanredef}
\begin{eqnarray}
\textrm{Galilean\ redefinition}:\hskip 0truecm &&E_\mu{}^0 = \omega\tau_\mu\,,\hskip .33truecm \ E_\mu{}^a = e_\mu{}^a\,,\hskip .33truecm E^\mu{}_0 = \frac{1}{\omega}\tau^\mu\,,\hskip .33truecm E^\mu{}_a= e^\mu{}_a\,,\\[.1truecm]
&&\Lambda^{0a} = \omega^{-1}\beta^a\,,\ \Lambda^{ab} = \lambda^{ab}\,.\label{Gal}
\end{eqnarray}
\end{subequations}
Here, the flat Lorentz index was split according to $A=(0,a)$ and we have introduced the boost-type parameter $\beta^a$.
We furthermore relabel $A_\mu=a_\mu$. These redefinitions lead to the following boost transformation rules (omitting the transport terms)
\begin{subequations}
\begin{eqnarray}\label{relGboosts1}
\delta \tau_\mu &=& \frac{1}{\omega^2} \beta_a e_\mu{}^a\,,\hskip 1truecm \delta \tau^\mu = -\beta^a e^\mu{}_a\,,\\[.1truecm]
\delta e_\mu{}^a &=& \beta^a\tau_\mu\,,\hskip 1truecm \delta e^{\mu a} = -\frac{1}{\omega^2}\tau^\mu\beta^a\,,\label{relGboosts2}
\end{eqnarray}
\end{subequations}
such that in the limit $\omega \to \infty$ we are left with the following non-trivial transformation  rules:
\begin{equation}
\delta_{\rm G} \tau^\mu = -\beta^a e^\mu{}_a\,,\hskip 1.5truecm \delta_{\rm G} e_\mu{}^a= \beta^a\tau_\mu\,.
\end{equation}
The subscript ${\rm G}$ serves as a reminder that these are the boosts after the contraction to the Galilei symmetry (up to transport terms). We note that, after taking the limit, the boosts are two-step nilpotent on $\tau^\mu$ and $e_\mu{}^a$ similar to~\eqref{eq:nilG}.

Substituting the Galilean redefinitions \eqref{Galileanredef}  into the relativistic Lagrangian \eqref{eq:realvecrel} we obtain the following redefined Lagrangian\footnote{We have ignored an overall factor of $\omega$ due to the fact that $E^{-1} = \omega^{-1} e^{-1}$ which can be absorbed by making an additional redefinition of the Maxwell field.}
\begin{equation}\label{nrexpansion}
  e^{-1} \mathcal{L}_{\rm rel} = +\frac{1}{2\omega^2}\tau^\mu\tau^\nu f_{\mu a}f_{\nu}{}^a - \frac14\, f_{ab} f^{ab} \,,
\end{equation}
where $e = \det (\tau_\mu, e_\mu{}^a)$ and  $f_{ab} = e_a{}^\mu e_b{}^\nu f_{\mu\nu}$ with $f_{\mu\nu} = \partial_\mu a_\nu - \partial_\nu a_\mu$.
Note that this is still a relativistic Lagrangian that is invariant under the relativistic boost transformations given in eqs.~\eqref{relGboosts1} and \eqref{relGboosts2}.

Taking the limit $\omega \to \infty$, we  obtain
\begin{equation}\label{mM}
e^{-1}\mathcal{L}_{\rm mG-Maxwell} =-\frac14\, e_a{}^\mu e^{a\rho} e_b{}^\nu e^{b\sigma} f_{\mu\nu} f_{\rho\sigma}\,,
\end{equation}
that is invariant under the following boost transformations:\footnote{We do not give the general coordinate transformations and the spatial rotations since these are manifest.}
\begin{equation}
\delta_{\rm G} e_\mu{}^a = \beta^a\tau_\mu\,. 
\end{equation}

Decomposing also the curved indices according to $\mu = (t,i)$
and imposing the following gauge fixing conditions corresponding to  a flat spacetime
\begin{equation}\label{gaugefixing}
\tau_\mu = \delta_\mu{}^t\,,\hskip 1truecm e_t{}^a = 0\,,\hskip 1truecm e_i{}^a = \delta_i{}^a\,,
\end{equation}
we obtain the following flat spacetime magnetic Galilean Maxwell Lagrangian:
\begin{equation}
  \label{eq:realvecnonrel2}
   \mathcal{L}_{\rm mG-Maxwell, flat} = - \frac14\, f_{ab} f^{ab} \,.
\end{equation}
The boost symmetries of this Lagrangian are the residual transformations of the gauge fixing  conditions \eqref{gaugefixing}:
\begin{equation}
\partial_t\xi^a + \beta^a = 0\,,
\end{equation}
which gives a transport term with $\xi^a = -t\,\beta^a$ or (note that $a_0$ is absent in the Lagrangian.)
\begin{equation}
\delta_{\rm G} a_b = -t\, \beta^c \partial_c a_b\hskip .3truecm \longrightarrow \hskip .3truecm \delta_{\rm G} f_{ab} = -t\, \beta^c\partial_c f_{ab}\,.
\end{equation}

Due to the absence of the field $a_0$ in the flat spacetime Lagrangian \eqref{eq:realvecnonrel2} there is an emergent
St\"uckelberg symmetry  $\delta a_0(x) = \rho(x)$ while the
 field equation corresponding to  $a_0$ does not follow  from
the non-relativistic Lagrangian \eqref{eq:realvecnonrel2}. The
situation is very similar to what happens when taking the limit of
Neveu--Schwarz gravity where the Poisson equation of the Newton
potential is missing \cite{Bergshoeff:2021bmc}. The single missing equation
of motion in this case is the equation $\vec{\nabla} \cdot \vec{E}=0$ which is obtained  by taking the limit of the relativistic
equations of motion. Denoting the missing equation of motion with $M$ and denoting the set of three equations of motion that do follow from the non-relativistic Lagrangean \eqref{eq:realvecnonrel2}  with NR, the complete set of non-relativistic equations of
motion form a reducible but indecomposable representation under
Galilean boosts which means that under boosts the missing equation of motion $M$  transforms to the  equations of motion  NR
but not the other way around. In other words, we have the following chain:
\begin{equation}
{\rm M} \longrightarrow {\rm NR} \longrightarrow 0\,.
\end{equation}

The flat spacetime case discussed here is precisely the  magnetic Galilean Maxwell theory discussed in \cite{Bellac}.
It is called magnetic since in the  limit the magnetic field dominates over the electric field:
\begin{equation}\label{redef2}
F_{0a} = 
\frac{1}{\omega}f_{0a}\,,\hskip 1.5truecm F_{ab} =  f_{ab}
\end{equation}
We have shown  how this theory can be coupled to Galilean gravity. The authors of \cite{Bellac} also discuss a second so-called electric limit where the electric field dominates the magnetic field. This case is not included in our discussion.

\subsection{Electric Carrollian Maxwell}

In the Carrollian case we redefine the gravitational fields as follows \cite{Bergshoeff:2017btm}:
\begin{subequations}
\label{Carrollianredef} 
\begin{eqnarray}
\textrm{Carrollian\ redefinition}:\hskip 0truecm && E_\mu{}^0 = \omega^{-1}\tau_\mu\,,\ E_\mu{}^a = e_\mu{}^a\,, E^\mu{}_0 = \omega \tau^\mu\,, E^\mu{}_A = e^\mu{}_a\,,\\[.1truecm]
&& \Lambda^{0a} = \omega^{-1}\beta^a\,,\ \Lambda^{ab} = \beta^{ab}\,.\label{Car}
\end{eqnarray}
\end{subequations}
Combining this Carrollian redefinition  with the gauge field redefinition
\begin{equation}
A_\mu = \omega^{-1} a_\mu
\end{equation}
we obtain the following Lagrangian:
\begin{equation}\label{expansion}
  e^{-1} \mathcal{L}_{\rm rel} = -\frac{1}{2}\tau^\mu\tau^\nu f_{\mu a}f_{\nu}{}^a - \frac14 \omega^{-2}\, f_{ab} f^{ab} \,,
\end{equation}
which, after taking the limit $\omega \to \infty$ leads to the following electric Carrollian Maxwell Lagrangian:
\begin{equation}\label{expansion2}
  e^{-1} \mathcal{L}_{\rm eC-Maxwell} = -\frac{1}{2}\tau^\mu\tau^\nu e^\rho{}_a e^{\sigma a}f_{\mu \rho}f_{\nu \sigma}\,.
\end{equation}
Ignoring transport terms, this Lagrangian is invariant under the following Carroll boost transformations with parameter $\beta_a$:
\begin{equation}
\delta_{\rm C}\tau_\mu = \beta_a e_\mu{}^a\,,\ \ \ \ \delta_{\rm C} \tau^\mu = 0\,,\ \ \ \ \delta_{\rm C} e_\mu{}^a = 0\,,\ \ \ \ \delta_{\rm C} e^\mu{}_a = -\beta_a \tau^\mu\,.
\end{equation}
Imposing the same gauge-fixing conditions \eqref{gaugefixing} as in the Galilean case, we obtain the following flat spacetime Lagrangian~\cite{Duval:2014uoa} (see also \cite{Henneaux:2021yzg,Islam:2023rnc})
\begin{equation}
   \mathcal{L}_{\rm eC-Maxwell, flat} = -\frac{1}{2} f_{0 a}f_0{}^a\,.
\end{equation}
Under Carroll boost transformations, this flat spacetime  Lagrangian only transforms with a transport term under time translations with parameter $\xi^t$ given by $\xi^t = -\beta^a x_a$.

Although the Lagrangian \eqref{expansion2} only depends on the electric field, there seems to be no emerging St\"uckelberg symmetry in this case.

\subsection{ Electric  Galilean Maxwell}

As an alternative to the seed Lagrangian approach, we  now consider a  second Galilean  limit that involves on the one hand the same Galilean redefinition \eqref{Galileanredef} of the gravitational fields but on the other hand makes use of the following different gauge field redefinition:
\begin{equation}
A_\mu = \omega\, a_\mu\,.
\end{equation}
This leads to the following Lagrangian:
\begin{equation}\label{electricGal}
  e^{-1} \mathcal{L}_{\rm rel} = +\frac{1}{2}\tau^\mu\tau^\nu  f_{\mu a}f_{\nu}{}^a - \frac14 \omega^2\, f_{ab} f^{ab} \,,
\end{equation}
with $f_{\mu\nu} = \partial_\mu a_\nu - \partial_\nu a_\mu$.
Applying the  quadratic divergence trick \eqref{quadratic} we obtain after taking the limit $\omega \to \infty$ the following electric Galilean Maxwell Lagrangian:
\begin{equation}\label{electricGal2}
  e^{-1} \mathcal{L}_{\rm eG-Maxwell} = +\frac{1}{2}\tau^\mu\tau^\nu e_a{}^\rho e^{a\sigma} f_{\mu \rho}f_{\nu\sigma} + \frac12 \chi^{ab}\, f_{ab}\,,
\end{equation}
Note that $f_{ab} = e_a{}^\mu e_b{}^\nu F_{\mu\nu}$ is invariant under boosts because $e_a{}^\mu$ is boost-invariant. The first term in the above Lagrangian transforms  under boosts as:
\begin{equation}
+ \beta^bf_{ab}\tau^\nu e^{a\sigma} f_{\nu\sigma}
\end{equation}
which is canceled by the following boost transformation of the Lagrange multiplier:
\begin{equation}\label{boostlmpl}
\delta_{\rm G} \chi^{ab} = + 2 \beta^{[a} \tau^\mu e^{b]\nu} f_{\mu\nu}\,.
\end{equation}
Alternatively, the transformation rule of the Lagrange multiplier $\chi^{ab}$ can  be obtained by first determining its transformation rule before taking the limit where the Lagrange multiplier can be solved for as $\chi^{ab} = -\omega^2 f^{ab}$ or
\begin{equation}
\delta \chi^{ab} = -\omega^2 \delta f^{ab} =  -\omega^2\delta \big(e^{\mu [a} e^{\nu b]} f_{\mu\nu}\big)\,.
\end{equation}
Substituting the boost transformation of $e^{\mu a}$ before taking the limit as  given in \eqref{relGboosts2} and next taking the limit $\omega \to \infty$ one obtains precisely the transformation rule \eqref{boostlmpl}.

One may verify that  the commutator of two boost transformations, with parameters $\beta^{a}_1, \beta^{a}_2$, on the Lagrange multiplier $\chi^{ab}$ closes up to a trivial symmetry with parameter $\sigma^{ab} = -\sigma^{ba}$ given by
\begin{equation}
\delta_{\rm G} \chi^{ab} = \sigma^{c[a} f_c{}^{b]}
\hskip .5truecm \textrm{with} \hskip .5truecm \sigma^{ab} = 4 \beta_2^{[a}\beta_1^{b]}\,.
\end{equation}

For a flat  spacetime the Lagrangian \eqref{electricGal2} reduces to
\begin{equation}\label{electricGal3}
   \mathcal{L}_{\rm eG-Maxwell,flat} = +\frac{1}{2} f_{0 a}f_{0}{}^a + \frac12 \chi^{ab}\, f_{ab}\,,
\end{equation}
Under boosts all fields now transform with a transport term which adds up to a total derivative. Besides these transport terms  we also have
\begin{equation}
\delta_{\rm G} a_0 = -\beta^b a_b\ \ \rightarrow\  \ \delta_{\rm G} f_{0a} = -\beta^b f_{ba}\,,\hskip .5truecm {\rm and} \hskip .5truecm \delta_{\rm G} \chi^{ab} = +2 \beta^{[a}f_0{}^{b]}
\end{equation}
such that the boost transformation of the first term in~\eqref{electricGal3} cancels against the boost transformation of the second term.
Note that the first term in this Lagrangian  is precisely the   electric Carrollian  Maxwell seed Lagrangian \eqref{expansion2}. The second  Lagrange multiplier term in \eqref{electricGal3}  is needed to compensate for the non-Galilean boost invariance of the leading $f_{0a}f_0{}^{a}$ term whose boost variation under Galilean boosts is proportional to $f_{ab}$. In contrast,  $f_{0a}$ is already invariant under Carrollian  boosts and, therefore, the corresponding electric Carrollian Maxwell seed Lagrangian \eqref{expansion2} is given by the  $f_{0a}f_0{}^{a}$ term only.

\subsection{ Magnetic  Carrollian Maxwell}

Substituting the Carrollian redefinitions \eqref{Carrollianredef} of the gravitational fields into the relativistic Lagrangian \eqref{eq:realvecrel}
and  relabeling $A_\mu = a_\mu$ we obtain the following Lagrangian:
\begin{equation}\label{nnrexpansion}
  e^{-1} \mathcal{L}_{\rm rel} = -\frac{1}{2}\omega^2\tau^\mu\tau^\nu f_{\mu a}f_{\nu}{}^a - \frac14\, f_{ab} f^{ab} \,.
\end{equation}
Applying the quadratic divergence trick \eqref{quadratic} thereby introducing a Lagrange multiplier $\chi^a$  and taking the limit $\omega \to \infty$, we obtain the following magnetic  Carrollian Maxwell Lagrangian:
\begin{equation}\label{nnrexpansion2}
  e^{-1} \mathcal{L}_{\rm mC-Maxwell} = \chi^a \tau^\nu f_{\nu a} - \frac14\, f_{ab} f^{ab} \,,
\end{equation}
which in flat spacetime yields the Lagrangian
\begin{equation}\label{nnrexpansion2flat}
 \mathcal{L}_{\rm mC-Maxwell, flat} = \chi^a  f_{0 a} - \frac14\, f_{ab} f^{ab} \,,
\end{equation}
We identify the second term as the magnetic Galilean Maxwell seed Lagrangian  \eqref{mM}.
This alternative seed method of constructing new Galilean or Carrollian  Maxwell theories is based
 on the following transformation property of the Maxwell curvature tensor under Galilean and Carrollian boosts in curved spacetime:
\begin{subequations}
\begin{eqnarray}
&&\textrm{Galilean \ boosts}:\hskip .8truecm f_{0a} \ \ \rightarrow \ \ f_{ab} \ \ \rightarrow\ \  0\,,\label{property}\\[.1truecm]
&&\textrm{Carrollian \ boosts}:\hskip .5truecm f_{ab} \ \ \rightarrow \ \ f_{0a} \ \ \rightarrow \ \ 0\,.\label{property2}
\end{eqnarray}
\end{subequations}
Starting from the magnetic Galilean Maxwell Lagrangian \eqref{mM}, we see that this Lagrangian is indeed invariant, due to property \eqref{property}, under Galilean boosts but that, due to property \eqref{property2},  the same Lagrangian transforms under Carrollian boosts to a term proportional to $f_{0a}$. Since, according  to property \eqref{property2}, $f_{0a}$ itself is invariant under Carrollian boosts, such a term can always be cancelled by adding a  term $\chi^{a}f_{0a}$ to the Lagrangian where $\chi^{a}$ is a Lagrange multiplier that transforms under Carrolian boosts such that the Lagrangian is invariant. This leads to the magnetic Carrollian Maxwell Lagrangian \eqref{nnrexpansion2}
where the Lagrange multiplier $\chi^{a}$ transforms under Carrolian boosts (ignoring the standard transport term) as follows:
\begin{equation}
\delta_{\rm C} \chi^{a} = \beta_b\, f^{ab}\,.
\end{equation}

\subsection{Yang--Mills}

The results obtained in this section on Maxwell  can be easily generalized to the Yang--Mills case.
The only place where one has to be careful is when we perform the Maxwell gauge field  redefinition $A_\mu = \omega^{\pm 1} a_\mu$  and make use of the fact that the Lagrangian is quadratic in the fields. The same overall scaling of the Lagrangian can be obtained in the Yang--Mills case by making an additional redefinition of the Yang--Mills coupling constant $G$
\begin{equation}
G =  g/\omega^{\pm 1}\,,
\end{equation}
such that
\begin{equation}
F^I_{\mu\nu} = \omega^{\pm 1} f^I_{\mu\nu} \hskip .5truecm\textrm{with}\hskip .5truecm f^I_{\mu\nu} = 2\partial_{[\mu} a^I_{\nu]} + g
f^I{}_{JK} a_\mu^J a_\nu^K\,.
\end{equation}

We note that Carrollian limits of abelian and non-abelian gauge theories have been studied in~\cite{Bagchi:2016bcd} and more recently in~\cite{Islam:2023rnc} where several models have been obtained by a different method. Some of the models obtained there agree with our results.

\subsection{\texorpdfstring{$p$-forms}{p-forms}}

The analysis above can be repeated for an (abelian) anti-symmetric $p$-form potential $A_{\mu_1\ldots \mu_p}$ with relativistic field strength $F_{\mu_1\ldots \mu_{p+1}} = (p+1) \partial_{[\mu_1} A_{\mu_2\ldots \mu_{p+1}]}$. The relativistic Lagrangian in flat space is
\begin{align}
\mathcal{L}_p = -\frac{1}{2(p+1)!} F_{\mu_1\ldots \mu_{p+1}} F^{\mu_1\ldots \mu_{p+1}}\,.
\end{align}
This theory can be coupled to gravity in the standard way, but we shall not display this coupling explicitly and content ourselves with the final flat space expressions.

Since the analysis is parallel to the one for the Maxwell field with $p=1$ we only state the final results.
The magnetic Galilei limit is
\begin{align}
\mathcal{L}_{\text{mG-$p$, flat}} = -\frac1{2(p+1)!} f_{a_1\ldots a_{p+1}} f^{a_1\ldots a_{p+1}}\,,
\end{align}
while the electric one becomes
\begin{align}
\mathcal{L}_{\text{eG-$p$, flat}} = \frac1{2 p!} f_{0 a_1\ldots a_{p}} f_0{}^{a_1\ldots a_{p}} + \frac1{(p+1)!} \chi^{a_1\ldots a_{p+1}} f_{a_1\ldots a_{p+1}} \,.
\end{align}

The Carrollian limits are for the electric case
\begin{align}
\mathcal{L}_{\text{eC-$p$, flat}} = -\frac1{2 p!} f_{0 a_1\ldots a_{p}} f^{0a_1\ldots a_{p}} \,,
\end{align}
while the magnetic one is
\begin{align}
\mathcal{L}_{\text{mC-$p$, flat}} = -\frac1{2(p+1)!} f_{a_1\ldots a_{p+1}} f^{a_1\ldots a_{p+1}} + \frac{1}{p!} \chi^{a_1\ldots a_{p}} f_{0 a_1\ldots a_p}\,.
\end{align}
For $p=0$, these theories reduce to the ones in section~\ref{sec:KG} in the case $m=0$.
For $p=1$, we re-obtain the Maxwell theories discussed in the beginning of this section.

\subsection*{Acknowledgements}
The authors would like to thank Glenn Barnich, Jos\'e Figueroa-O'Farrill and Alfredo Per\'ez for discussion. We also benefitted from conversations with the participants of the workshop Carroll@Vienna 2022.
The work of JG has been supported in part by MINECO FPA2016-76005-C2-1-P
and PID2019-105614GB-C21 and from the State Agency for Research of the
Spanish Ministry of Science and Innovation through the Unit of Excellence
Maria de Maeztu 2020-203 award to the Institute of Cosmos Sciences
(CEX2019-000918-M).


\appendix

\section{Massive Galilei as a seed}
\label{app:massG}

In this appendix, we consider the massive Galilei particle whose symmetry algebra is the Bargmann algebra that is the central extension of the Galilei algebra~\eqref{eq:Gal} by a central generator $M$ appearing in the commutator
\begin{align}
\label{eq:a1}
[ B_a , P_b ] =\delta_{ab} M\,.
\end{align}
The canonical particle Lagrangian for mass $m>0$ is given by
\begin{align}
\label{eq:massG}
L_{\rm B} = - E \dot{t} + \vec{p} \cdot \dot{\vec{x}} -\frac{e}{2} \left( 2m E -\vec{p}^{\, 2}\right)\,,
\end{align}
showing the standard non-relativistic energy of a massive particle.

The Bargmann boosts now act as 
\begin{subequations}
\label{eq:Bargb}
\begin{align}
\delta_{\rm B} x^a &= \beta^a \, t \,,& \delta_{\rm B} p^a &= m \beta^a\,,\\
\delta_{\rm B} t &= 0\,,& \delta_{\rm B} E &= \vec{\beta}\cdot \vec{p}
\end{align}
\end{subequations}
along with $\delta_{\rm B} e =0$.  Comparing these transformations with~\eqref{eq:TRG0} we see that the transformation of $p^a$ has changed and that now there is a three-step nilpotency
\begin{align}
\label{eq:nilB}
E \longrightarrow p^a \longrightarrow m \longrightarrow 0 \,.
\end{align}
The symplectic term is no longer strictly invariant but transforms into a total derivative. This gives rise to modified Noether charges, implying a central in their algebra as shown in~\eqref{eq:a1}.

Treating the Lagrangian~\eqref{eq:massG} as a seed, we can apply the Carroll transformations~\eqref{eq:TRC0} (without any deformation for general $D$) to it and obtain
\begin{align}
\delta_{\rm C} L_{\rm B} = e \vec{\beta} \cdot \vec{p} E = \delta_{\rm C} \left( \chi E \right)\,,
\end{align}
if the thus introduced Lagrange multiplier $\chi$ transforms as
\begin{align}
\delta_{\rm C} \chi = e \vec{\beta} \cdot \vec{p}
\end{align}
as before. Here, we relied again upon the two-step nilpotency of the Carroll transformations~\eqref{eq:TRC0} since there is no central extension of the Carroll algebra for $D>2$.

Therefore we conclude that the Lagrangian
\begin{align}
L_{\rm C} = - E \dot{t} + \vec{p} \cdot \dot{\vec{x}} -\frac{e}{2} \left( 2m E -\vec{p}^{\, 2}\right) - \chi E
\end{align}
is invariant under Carroll transformations. Performing a canonical analysis one finds no degrees of freedom. This can be traced back to the constraints now including $E= \frac{\vec{p}^{\, 2}}{2m} = 0$, implying $\vec{p}=0$ which is stronger than the condition encountered for the magnetic massive Carroll particle in section~\ref{sec:MCP}.

We also consider the opposite direction, i.e., starting from a Carroll-invariant Lagrangian and aim to make it Bargmann-invariant. For illustration, we start from the time-like massive Carroll theory (see~\eqref{eq:EC0})
\begin{align}
L_{\rm eC} = - E \dot{t} + \vec{p} \cdot \dot{\vec{x}}
-\frac{e}{2} \left( -E^2+ m^2 \right)
\end{align}
and consider the action of the Bargmann boosts~\eqref{eq:Bargb} on it (up to total derivatives):
\begin{align}
\delta_{\rm B} L_{\rm eC} =  e E \delta_{\rm B} E = e E \beta^a p_a\,.
\end{align}
Since $p_a$ is not invariant under Bargmann boosts, we cannot make this Lagrangian invariant by adding a single Lagrange multiplier $\chi^a$ as in section~\ref{sec:EGP}. This is also reflected in the three-step nilpotency~\eqref{eq:nilB}. The resolution is to introduce two Lagrange multiplier variables $\chi^a$ and $\varphi$ in the form
\begin{align}
L_{\rm B} = - E \dot{t} + \vec{p} \cdot \dot{\vec{x}}
-\frac{e}{2} \left( -E^2+ m^2 \right) - \chi^a p_a - m \varphi
\end{align}
with Bargmann boost transformations
\begin{align}
\delta_{\rm B} \chi^a =  e E \beta^a \,,\quad\quad
\delta_{\rm B} \varphi &= -\beta^a \chi_a\,.
\end{align}
This system is invariant under Bargmann boosts but has no physical degrees of freedom. The variables now form a five-step nilpotent chain under boosts
\begin{align}
\varphi \longrightarrow \chi^a \longrightarrow E \longrightarrow p^a \longrightarrow m \longrightarrow 0\,.
\end{align}
This method of resolving the invariance of a Lagrangian with a three-step nilpotency can be generalised to $n$-step chains in a straight-forward manner.

\medskip

For $1+1$ dimensions, the Carroll algebra also admits a central extension and the isomorphism between unextended Galilei and Carroll extends to this case.
The map~\eqref{eq:map} can be applied to the Bargmann Lagrangian~\eqref{eq:massG} in $1+1$ dimensions to yield
\begin{align}
L_{\rm extC} = -E\dot{t} + p\dot{x} +\frac{e}{2} (2mp + E^2)
\end{align}
which reduces to
\begin{align}
L_{\rm extC} =  \frac{m\dot{t}^2}{2\dot{x}}\,.
\end{align}
on a physical branch after integrating out the auxiliary variables.

\end{document}